\documentclass[10pt]{iopart}

		\expandafter\let\csname equation*\endcsname\relax
		\expandafter\let\csname endequation*\endcsname\relax

	\usepackage{bm}					
	\usepackage{amsmath}
	\usepackage{mathtools}
	\usepackage{amssymb}

	\allowdisplaybreaks				
	
	\usepackage{float}				
	\usepackage{graphicx}			

\begin{document}

\title{Pulse amplification in a closed loop $\Lambda$ system with permanent dipole moments}

\author{Nilamoni Daloi$^1$\footnote{email id: nilamoni@iitg.ac.in}, Partha Das$^1$\footnote{email id: partha.2015@iitg.ac.in}, and Tarak Nath Dey$^1$\footnote{email id: tarak.dey@iitg.ac.in}}
\address{$^1$Indian Institute of Technology Guwahati, Guwahati, Assam, India}

\vspace{10pt}
\begin{indented}
\item[]\date{\today}
\end{indented}

\begin{abstract}
Propagation of a weak Gaussian probe pulse through a closed loop $\Lambda$ system with permanent dipole moments (PDMs) is investigated in presence of a strong control field along with a third field. The presence of PDMs allows multi photon excitation, which are otherwise forbidden. The PDMs modify the Rabi frequencies of the probe, control, and the third field inside the medium which noticeably affects the propagation of probe pulse. The probe pulse is amplified during propagation with its Gaussian shape intact. Due to unprohibited two photon excitation it is possible to amplify a probe pulse whose frequency is twice of the control field's frequency, with the help of the third field.
\end{abstract}

%
%
%
\ioptwocol

\section{\label{sec:level1}Introduction}

In presence of a high intensity field, called the control field, an opaque medium can be rendered transparent for a low intensity probe field, within a narrow spectral range around the absorption line of the probe field. This phenomenon, first proposed by Harris \textit{et al.} in 1990 \cite{eit_1}, is called electromagnetically induced transparency (EIT) \cite{eit_2}. The narrow transparency window is accompanied by strong dispersion which can be used to produce ``slow light" \cite{slow_light_1,slow_light_2,slow_light_3}. By exploiting the atomic coherences, a light pulse can also be stored and retrieved \cite{storage_and_retrieval_1, storage_and_retrieval_2,storage_and_retrieval_3,storage_and_retrieval_4} in an EIT system. STIRAP \cite{stirap_1,stirap_2,stirap_3}, CPT \cite{eit_2}, and lasing without inversion \cite{LWI_1,LWI_2,LWI_3} are some popular phenomenons where atomic coherence has been exploited for practical applications. All experimental and theoretical studies of EIT and aforementioned phenomena have been mostly based on atoms. Recently there have been studies based on polar  molecules. Due to the presence of permanent dipole moments (PDM) in polar molecules, new phenomena emerges which are significantly different from the ones in absence of PDM. So far, there has been studies on the effects of PDMs on EIT \cite{eit_pdm_1,eit_pdm_2}, STIRAP \cite{stirap_pdm_1,stirap_pdm_2}, pulse propagation \cite{propagation_pdm_1}, higher harmonic generation \cite{higher_harmonic_generation_pdm_1}, population inversion \cite{pop_inversion_pdm_1} and optical bistability \cite{optical_bistability_pdm_1}. In this work, we study the propagation of a weak probe pulse through a closed loop three level $\Lambda$ system with PDMs in presence of a strong control field and a third field.

Molecules can be classified mainly into centrosymmetric and non centrosymmetric. Centrosymmetric molecules have a spatial inversion center while noncentrosymmentic molecules (e.g. all polar molecules) don't. Due to which, the energy eigenfunctions, $\psi$s for centrosymmentric molecules are parity eigenstates, meaning they are either even or odd functions. Whereas wave functions for non centrosymmetric molecules are neither even nor odd. Thus, the diagonal elements of the dipole moment matrix, $\mu_{jj}\propto\int_{-\infty}^\infty \psi_j\vec{r}\psi_jdr $, are zero for centrosymmmetric molecules and non zero for non centrosymmetric molecules. These diagonal matrix elements represent the permanent dipole moments as indicated in Fig. \ref{fig:level_system}. Also in case of centrosymmetric molecules $\mu_{jk}\neq 0$ only if $\psi_j$ and $\psi_k$ are of different parity i.e., if one of them is odd then the other must be even and vice versa, therefore a two photon transition between $|\psi\rangle_j\rightarrow|\psi\rangle_k$ is possible via an intermediate state $|\psi\rangle_i$ provided $|\psi\rangle_j$ and $|\psi\rangle_k$ have same parity, in which case they are single photon forbidden. Therefore, for centrosymmetric molecules the single photon allowed transitions are not two photon allowed and vice versa. But in case of noncentrosymmetric molecules, since $\psi$'s are neither odd nor even therefore both single and two photon excitation is possible for the same transition.

\section{\label{sec:level2}Theory}

\begin{figure}[!t]
\centering
\includegraphics[width=0.9\linewidth]{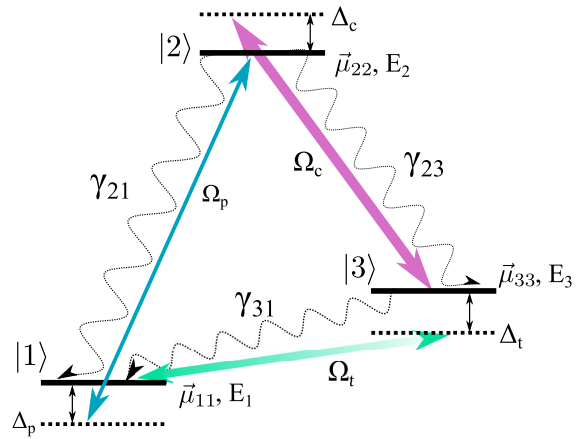}
\caption{\label{fig:level_system}Schematic diagram of a closed loop three level $\Lambda$-system with PDMs. Here, $|2\rangle$ is the excited state, $|3\rangle$ is an intermediate meta stable state and $|1\rangle$ the ground state, with energy set to zero. The molecular transitions, $|1\rangle \leftrightarrow|2\rangle$, $|3\rangle\leftrightarrow|2\rangle$, and $|1\rangle\leftrightarrow|3\rangle$ are coupled by a weak probe field, $\vec{E}_p$ with frequency, $\omega_p$, a strong control field, $\vec{E}_c$ with frequency, $\omega_c$, and a third field, $\vec{E}_t$ with frequency, $\omega_t$,  respectively. The population decay rate from state, $|i\rangle$ to $|j\rangle$ is denoted by $\gamma_{ij}$. The detunings, and Rabi frequencies of the fields are denoted by $\Delta_i$, and $\Omega_i$ respectively ($i = p,c,t $ represents probe, control, and the third field, respectively). The energies of $|i\rangle$ ($i = 1,2,3$) are denoted by $E_i$ and the PDMs are denoted by $\mu_{ii}$.}
\end{figure}

To study the effect of PDMs on pulse propagation, we consider a closed loop three level $\Lambda$ system with PDMs as given in Fig. \ref{fig:level_system}. In Fig. \ref{fig:level_system}, the molecular transitions $|1\rangle \leftrightarrow|2\rangle$, $|3\rangle\leftrightarrow|2\rangle$, and $|1\rangle\leftrightarrow|3\rangle$ are coupled by a weak probe field, $\vec{E}_p$ with frequency, $\omega_p$, a strong control field, $\vec{E}_c$, with frequency, $\omega_c$, and a third field, $\vec{E}_t$ with frequency, $\omega_t$, respectively. The electric fields are considered to be propagating along $z$ direction and are defined as:

\begin{equation}
\vec{E} = \sum_{j=p,c,t}\hat{e}_j\mathcal{E}_j f_j(t)\cos(\omega_j t)\label{eq:electric_field}
\end{equation}		
where $\hat{e}_j$ are the unit polarization vectors, $\mathcal{E}_j$ are the field amplitudes, $f_j(t)$ are the slowly varying envelope functions, and $\omega_j$ are the field carrier frequencies of the fields. The index $j= p, c, t$ represents
probe, control, and the third field, respectively. Let $|\psi(t)\rangle = \sum_i^3a_j(t)|j\rangle$ be a general state for the  $\Lambda$ system in Fig. \ref{fig:level_system} such that the Schr\"{o}dinger equation for the $\Lambda$ system, in terms of the probability amplitudes $a_j$, can be written as:	

\begin{equation}
i\frac{\partial \textbf{a}(t)}{\partial t} = \hbar^{-1}\bigg[\textbf{E} -\bm{\mu} . \vec{E}(t)\bigg]\textbf{a}(t),\label{eq:2}
\end{equation}
where, $(\textbf{a}(t))_j = a_j(t)$, $(\textbf{E})_{jk} = E_{j}\delta_{jk}$, and $\bm{(\mu})_{jk} = \vec{\mu}_{jk} = \langle j|\bm{\mu}|k\rangle$. To write the Hamiltonian in a time independent form, the following transformation \cite{eit_pdm_1,eit_pdm_2,unitary_transformation} is used:

\begin{equation}
\textbf{a(t)} = \textbf{T}\textbf{b}(t),\label{eq:transformation_equation}
\end{equation}
where
\begin{equation}
T_{jk} = \delta_{jk}\exp[-i\frac{ E_{j}}{\hbar}(t-t_0)]\exp[i\beta_{jk}],\label{eq:transformation_matrix}
\end{equation}
and
\begin{equation}
\beta_{jk} = \frac{\vec{\mu}_{jk}}{\hbar} .\int_{0}^t\vec{E}(t^\prime)dt^\prime,\label{eq:5}
\end{equation}
Using Eqs. [\eqref{eq:transformation_matrix}, \eqref{eq:5}] and substituting Eq. \eqref{eq:transformation_equation} into Eq. \eqref{eq:2} gives:
\begin{equation}
i\frac{\partial \textbf{b}(t)}{\partial t} = \textbf{H}^b(t)\textbf{b}(t),\label{eq:6} 
\end{equation}
with
\begin{equation}
\textbf{H}_{jk}^b(t) =-\frac{\vec{\mu}_{jk}.\vec{E}(t)}{\hbar}\exp[-\frac{iE_{kj}}{\hbar}t]\exp[i(\beta_{kk}-\beta_{jj})],\label{eq:Hamiltonian_1}
\end{equation}
where, $\textbf{H}_{jk}^b(t) = 0,\;( \forall\; j=k)$ and $E_{k,j}=E_k - E_j$. Putting \eqref{eq:5} in \eqref{eq:Hamiltonian_1}, expanding the cosine functions
in terms of exponentials, and using $\exp(iz\sin x)=\sum_{n=-\infty}^\infty J_n(z)\exp(inx)$ [$J_n(z)$ is the Bessel
function of first kind of order $n$ ($n\in\mathbb{Z}$)], gives:	

\begin{align}
\textbf{H}_{jk}^b(t)&=-\sum_{\substack{n_i=-\infty}}^{\infty}\left(\sum_i\frac{n_i\vec{\mu}_{jk}.\hat{e}_i\mathcal{E}_i}{\hbar z_{kj}^i}  \right)\prod_i J_{n_i}\left(z_{kj}^if_i(t)\right)\notag\\
&\times\exp\left[-i\left(\frac{E_{kj}}{\hbar}-\sum_{n_i}n_i\omega_i\right) t\right],\quad\left(i = p,c,t\right)\label{eq:Hamiltonian_2}
\end{align}
Now applying RWA (Rotating wave approximation), the far-off-resonant terms in the exponentials are neglected. It is also assumed that the probe 
transition involves the absorption of $n_p$ probe photons, and zero photons of control and the third field. Same rule applies for the transitions corresponding to control and the third field. Finally the Hamiltonian for the system becomes:

\begin{align}
\label{eq:Hamiltonian_3}
\textbf{H}^b(t) &=-\bigg(\tilde{\Omega}_pe^{i\Delta_pt}|1\rangle\langle 2| +
\tilde{\Omega}_te^{i\Delta_t t}|1\rangle\langle 3|\notag\\
& + \tilde{\Omega}_c^*e^{-i\Delta_c t}|2\rangle\langle 3|\bigg) + \text{h.c.,}
\end{align}
where, $\Delta_p = n_pw_p - E_{21}/\hbar$, $\Delta_c = n_cw_c - E_{23}/\hbar$, and $\Delta_t = n_tw_t - E_{31}/\hbar$ are the respective field detunings, and $\tilde{\Omega}_i$ ($i = p,c,t$) are the effective Rabi frequencies of probe (p), control (c) and the third (t) field,  respectively, in presence of PDMs. The effective Rabi frequencies of the three fields can be expressed as:

\begin{subequations}
\label{eq:efffective_rabi_frequencies}
\begin{align}
\tilde{\Omega}_p &=\frac{ n_p\Omega_p J_{n_p}(z_{21}^pf_p(t)) J_0(z_{21}^cf_c(t)) J_0(z_{21}^tf_t(t))}{z^p_{21}},\label{eq:efffective_probe_rabi_frequency}\\
\tilde{\Omega}_c &= \frac{n_c\Omega_c J_0(z^p_{23}f_p(t)) J_{n_c}(z^c_{23}f_c(t)) J_0(z^t_{23}f_t(t))}{z^c_{23}},\label{eq:efffective_control_rabi_frequency}\\
\tilde{\Omega}_t &= \frac{n_t\Omega_m J_0(z^p_{31}f_p(t)) J_0(z^c_{31}f_c(t)) J_{n_t}(z^t_{31}f_t(t))}{z^t_{31}}, \label{eq:efffective_third_rabi_frequency}
\end{align}
\end{subequations}
with

\begin{align}
&\Omega_{p} = \frac{\vec{\mu}_{12}.\hat{e}_p\mathcal{E}_p}{\hbar},\;\Omega_{c} = \frac{\vec{\mu}_{32}.\hat{e}_c\mathcal{E}_c  }{\hbar},\; \Omega_{t} = \frac{\vec{\mu}_{13}.\hat{e}_t\mathcal{E}_t}{\hbar},
\notag\\
&\text{and}\;z^\alpha_{jk} = \frac{\vec{d}_{jk}.\hat{e}_\alpha\mathcal{E}_\alpha}{\hbar w_\alpha}\;(\alpha = p,c,t)\label{eq:z_ij}.
\end{align}
In Eq. \eqref{eq:z_ij}, $\vec{d}_{jk} = \vec{\mu}_{jj} - \vec{\mu}_{kk}\;(j,k = 1,2,3)$. In Eqns. \eqref{eq:efffective_probe_rabi_frequency}, \eqref{eq:efffective_control_rabi_frequency}, and \eqref{eq:efffective_third_rabi_frequency} the indices $n_p$, $n_c$, $n_t$ determines whether the probe, control, and the third field transitions, respectively are one or multiphoton transitions. For example, $n_p = n_c = n_t = 1$ refers to all three transitions being one photon transitions. While, $n_p = n_c = n_t = 2$, would mean, all three transitions are two photon transitions. The molecular level system given in Fig. \ref{fig:level_system}, can be realized using the vibrational energy levels of HCN. The energies and PDMs of relevant HCN vibrational energy levels are listed in  table \ref{table:parameters}.

\begin{table}[H]
\begin{center}
\begin{tabular}{ c c c c }
 $|i\rangle$ & $\nu_1$, $\nu_2$, $l$, $\nu_3$ & $E_i$ (a.u.) & $\mu_{ii}$ (a.u.)		\\
\hline
\hline  
 $|1\rangle$ & $0,0,0,0$ & 0 		& 1.17061					 						\\

 $|2\rangle$ & $3,1,1,0$ & 0.047038 	& 1.181391										\\
  $|3\rangle$ & $0,1,1,0$ & 0.003269 	& 1.151803										\\
 \hline
\end{tabular}
\caption{\label{table:parameters} HCN$\rightarrow$HNC isomerization data. Here, a.u. means atomic unit.}
\end{center}
\end{table}

\noindent The transition dipole moments in atomic units (a.u.) are: $\mu_{12} =1.25906\times10^{-5}$, $\mu_{13} = 0.074363184$, $\mu_{32} = 3.93\times10^{-5}$. The above paramenters and data given in table \ref{table:parameters} are taken from HCN$\rightarrow$HNC isomerization data \cite{HCN_data_1, HCN_data_2}.

\subsection{Density matrix equations}
The dynamics of molecular energy state population and molecular coherences are governed by the following Liouville equation:

\begin{equation}
\frac{\partial \rho}{\partial t} = -i\hbar [\bm{H}^b(t), \rho] + \mathcal{L}_{\rho}.
\end{equation}
The Liouville operator $\mathcal{L}_{\rho}$, describes all incoherent processes and can be expressed as:

\begin{equation}
\mathcal{L}_{\rho} = -\sum_{i=2}^3 \sum_{\substack{j=1 ,\\ j\neq i}}^3 \frac{\gamma_{ij}}{2}\left(|i\rangle\langle i|\rho - 2 |j\rangle\langle j|\rho_{ii} + \rho |i\rangle\langle i|\right),
\end{equation}
To get rid of the exponential factors in Eq. \eqref{eq:Hamiltonian_3}, the following transformations are used:
$\rho_{12} = \rho_{12}e^{i\Delta_pt}, \rho_{23} = \rho_{23}e^{-i\Delta_ct}, \rho_{13} = \rho_{13}e^{i(\Delta_p - \Delta_c)t}\; $ and $ \rho_{jj} = 			\rho_{jj}\;(j = 1,2,3)$. The equations of motion for the molecular state populations and coherences of the three level $\Lambda$-system are then given as:
{\footnotesize
\begin{subequations}
\label{eq:density_matrix_equations}
\begin{align}
\dot{\rho}_{11} &= \gamma_{21}\rho_{22} + \gamma_{31}\rho_{33} + i\left(\tilde{\Omega}_p\rho_{21} - \tilde{\Omega}^*_p\rho_{12} + \tilde{\Omega}_t\rho_{31} - \tilde{\Omega}^*_t\rho_{13}\right),\\
\dot{\rho}_{22} &= -(\gamma_{21} + \gamma_{23})\rho_{22}-i\left(\tilde{\Omega}_p \rho_{21} - \tilde{\Omega}^*_p\rho_{12}  +\tilde{\Omega}_c \rho_{23}- \tilde{\Omega}^*_c\rho_{32} \right),\\
\dot{\rho}_{33} &=  \gamma_{23}\rho_{22}- \gamma_{31}\rho_{33} - i\left(\tilde{\Omega}^*_c\rho_{32} - \tilde{\Omega}_c\rho_{23} +\tilde{\Omega}_t\rho_{31} - \tilde{\Omega}^*_t\rho_{13}\right),\\
\dot{\rho}_{12} &= -\left(\Gamma_{21} + i \Delta_p\right)\rho_{12} - i\bigg[e^{i\delta t}\left(\tilde{\Omega}_c\rho_{13} - \tilde{\Omega}_t\rho_{32}\right)\notag\\ 
&+ \tilde{\Omega}_p\left(\rho_{11} - \rho_{22}\right)\bigg],\\
\dot{\rho}_{13} &= -\left(\Gamma_{31} + i\Delta_t\right)\rho_{13} - i \bigg[e^{-i\delta t}\left(\tilde{\Omega}^*_c\rho_{12} - \tilde{\Omega}_p\rho_{23}\right) \notag\\
&+ \tilde{\Omega}_t\left(\rho_{11} - \rho_{33}\right)\bigg],\\ 
\dot{\rho}_{23} &= -(\Gamma_{23} - i\Delta_c)\rho_{23} - i\bigg[ e^{i\delta t}\left( \tilde{\Omega}_t\rho_{21} - \tilde{\Omega}^*_p\rho_{13}\right) \notag\\
&+\tilde{\Omega}^*_c(\rho_{22} - \rho_{33})  \bigg],\\
\rho^*_{ij} &= \rho_{ji},\quad \text{and}\quad \sum^3_{j =1}\rho_{jj} = 1.
\end{align}
\end{subequations}
}
\noindent Here, $\delta = \left(\Delta_c + \Delta_t - \Delta_p\right)$ is the three photon detuning. The above equations are normalized with respect to $\gamma_{21}$ and solved using $5$\textsuperscript{th} order Runge Kutta method with initial conditions: $\rho_{11} = 1, \rho_{22} = \rho_{33}=0\; \forall z$. In Eq. \eqref{eq:density_matrix_equations}, the
overdots stand for time derivatives and ``*" denotes complex
conjugate. The decay rate of population form state $|i\rangle$ to $|j\rangle$ are denoted by $\gamma_{ij}$ and the decoherence rate of  $\rho_{12}$, $\rho_{23}$, $\rho_{13}$ are given as $\Gamma_{12} = (\gamma_{21} +\gamma_{23})/2$, $\Gamma_{23} = (\gamma_{21} +\gamma_{23}+ \gamma_{31})/2$, $\Gamma_{31} = \gamma_{31}/2$, respectively. The data for longitudinal and transverse relaxation rates pertaining to the vibrational levels given in table. \ref{table:parameters} are unavailable. However, we assume $\gamma_{21} = \gamma_{23} = \gamma_{31}= \gamma \approx 10^{12} Hz$, which is of the order of the reorientation rate of molecules
in solution \cite{decay_rate}.

\subsection{Propagation equations}
To study the spatiotemporal evolution of the probe pulse and the CW control field, the following Maxwell-Bloch equations are used:

\begin{subequations}
\label{eq:Maxwell_equations}
\begin{align}
\bigg(\frac{\partial}{\partial z} + \frac{1}{c}\frac{\partial}{\partial t}\bigg)\tilde{\Omega}_p(z,t) &=i\eta^\prime_p \rho_{21}(z,t), \label{eq:Maxwell_equations_1}\\
\bigg(\frac{\partial}{\partial z} + \frac{1}{c}\frac{\partial}{\partial t}\bigg)\tilde{\Omega}_c(z,t) &=i\eta^\prime_c \rho_{23}(z,t).\label{eq:Maxwell_equations_2}
\end{align}
\end{subequations}
\noindent Here $\eta^\prime_i$ $(i= p,c)$ are called the coupling constants of the respective fields:

\begin{align*}
\label{eq:coupling_constants}
\eta^\prime_p &= n_p\eta_p\left|\frac{J_{n_p}(z_{21}^pf_p(t)) J_0(z_{21}^cf_c(t)) J_0(z_{21}^tf_t(t))}{z^p_{21}}\right|^2,\\
\eta^\prime_c &= n_c \eta_c\left|\frac{J_0(z^p_{23}f_p(t)) J_{n_c}(z^c_{23}f_c(t)) J_0(z^t_{23}f_t(t))}{z^c_{23}}\right|^2,
\end{align*}
where $\eta_i = 3N\lambda^2_i\gamma/8\pi$, with $N$ being the number of molecules per unit volume, and $\lambda_i$, the wavelength of respective fields. To facilitate numerical integration of Eq. (\ref{eq:Maxwell_equations}), a frame moving at the speed of light in vacuum, $c$ is used. The necessary coordinate transformations for that are $\tau = t - z/c$, and $\zeta = z$. This allows for the round bracketed terms of Eq. (\ref{eq:Maxwell_equations}) to be replaced by partial derivatives with respect to the single independent variable $\zeta$. Since $\lambda_t>>\lambda_{p,c}$, propagation dynamics of $\vec{E}_t$ has been ignored.
%
%
\section{Results}
\begin{figure}[!b]
\begin{center}
\includegraphics[width=0.97\linewidth]{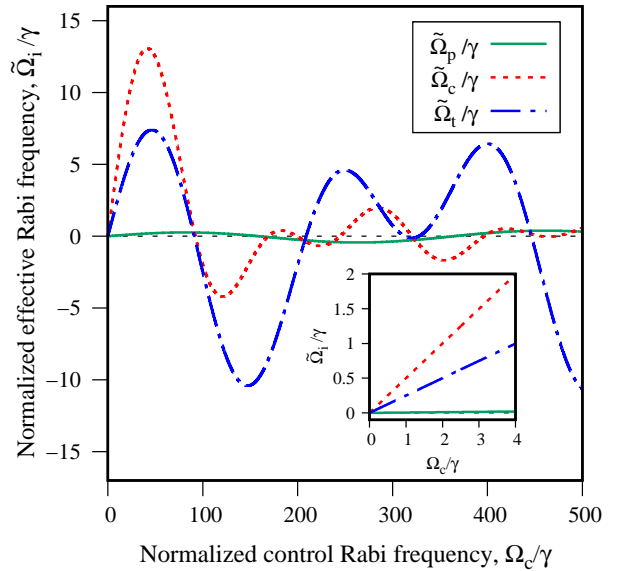}
\end{center}
\caption{\label{fig:effective_rabi_frequency_plot}$\tilde{\Omega}_i/\gamma$ vs $\Omega_c/\gamma$. Parameters used: $n_p=n_c=n_t = 1$ (meaning all three corresponding to probe, control and the third field are one photon transitions, respectively), $\Omega_p = 0.01\Omega_c$, $\Omega_t = 0.5\Omega_c$, $\Delta_p = \Delta_c = \Delta_t = 0 $.} 
\end{figure}
In presence of PDM, the usual Rabi frequencies, $\Omega_i$ gets modified as the effective Rabi frequencies, $\tilde{\Omega}_i$, inside the medium. Figure \ref{fig:effective_rabi_frequency_plot} shows the variation of effective Rabi frequencies of probe, control, and the third field with respect to the usual control Rabi frequency. In Fig. \ref{fig:effective_rabi_frequency_plot}, the effective Rabi frequencies show oscillatory behavior with respect to $\Omega_c/\gamma$ due to the presence of Bessel functions in the expressions of $\tilde{\Omega}_i$ as seen in Eq. \eqref{eq:efffective_rabi_frequencies}. However within the experimentally feasible value of $\Omega_c$, the variation of $\tilde{\Omega}_i$ with $\Omega_c$ is linear as shown in the inset plot of Fig. \ref{fig:effective_rabi_frequency_plot}. 

We shall investigate the propagation of a weak Gaussian probe pulse in presence of a continuous wave (CW) control and the third field, i.e.,
\begin{equation}
\label{eq:Gaussian_probe_profile}
f_p(\tau) = \exp[-\frac{(\tau-\tau_0)^2 }{2\sigma_0^2}],\;f_c(\tau) = 1,\; \text{and} \;f_t(\tau) = 1.
\end{equation}
\begin{figure}[!t]
\begin{center}
\includegraphics[width=\linewidth]{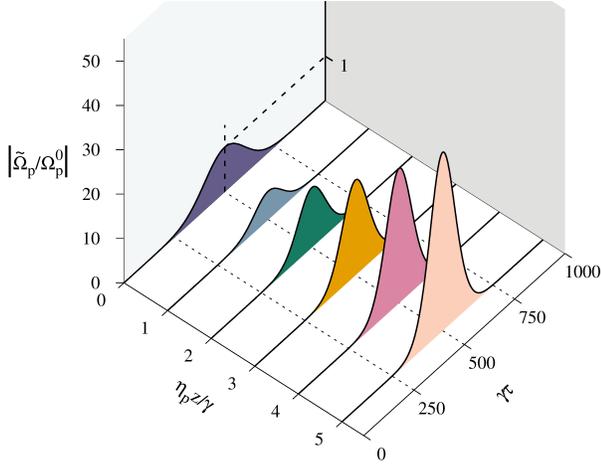}
\end{center}
\caption{\label{fig:effective_probe_rabi_frequency_at_different_z}$\left|\tilde{\Omega}_p/\Omega^0_p\right|$ vs $\gamma \tau$ at different normalized propagation length, $\eta_pz/\gamma$. Parameters used: $n_p= 1, n_c=n_t = 1$ (meaning all three transitions are one photon), $\Omega_p = 0.01\Omega_c$, $\Omega_t = 0.5\Omega_c$, $\Omega_c = 1\gamma$, $\Delta_p = \Delta_c = \Delta_t = 0 $, $\sigma_0 = 100/\gamma$, $\tau_0 = 500/\gamma$. Here, $\Omega^0_p$ denotes the usual probe Rabi frequency at $z = 0$. The right and left $z$ axes represent normalized probe field magnitude at $z = 0$ and $z > 0$, respectively.} 
\end{figure}
\begin{figure}[!t]
\begin{center}
\includegraphics[width=0.98\linewidth]{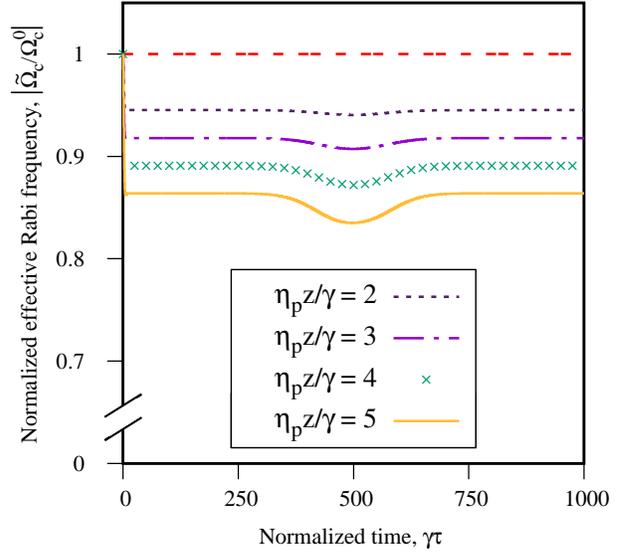}
\end{center}
\caption{\label{fig:effective_control_rabi_frequency_at_different_z}$\left|\tilde{\Omega}_c/\Omega^0_c\right|$ vs $\gamma \tau$ at different normalized propagation length, $\eta_pz/\gamma$. Parameters used are same as Fig. \ref{fig:effective_probe_rabi_frequency_at_different_z}. Here, $\Omega^0_c$ denotes the probe Rabi frequency at $z = 0$.} 
\end{figure}
\begin{figure}[!t]
\begin{center}
\includegraphics[width=0.98\linewidth]{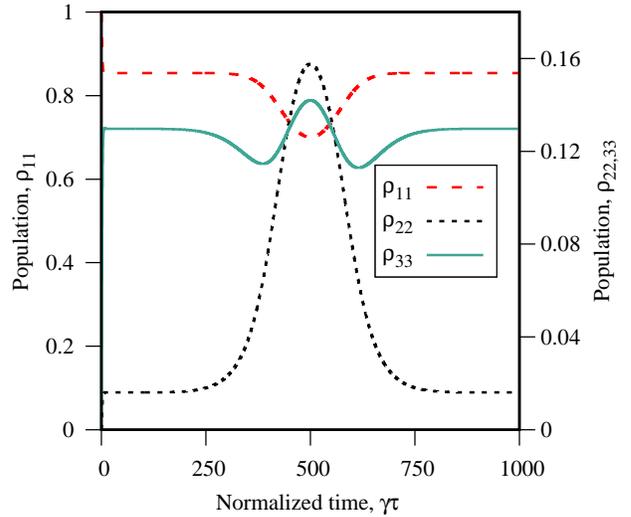}
\end{center}
\caption{\label{fig:population_distribution}Population, $\rho_{ii}$ vs $\gamma \tau$ at normalized propagation length, $\eta_pz/\gamma = 5$. Parameters used are same as Fig. \ref{fig:effective_probe_rabi_frequency_at_different_z}.} 
\end{figure}
\noindent Here, $\sigma_0$ and $\tau_0$ are the probe pulse width and delay, respectively at the entrance of the medium. We first consider the one photon excitation case, i.e., all three fields are considered to exciting one photon transitions ($n_p = n_c= n_t = 1$). Figure \ref{fig:effective_probe_rabi_frequency_at_different_z} shows the temporal profile of the Gaussian probe pulse at different normalized propagation length, $\eta_pz/\gamma$. In Fig. \ref{fig:effective_probe_rabi_frequency_at_different_z}, the probe pulse can be seen getting amplified without any broadening and delay. Figure \ref{fig:effective_control_rabi_frequency_at_different_z} shows the temporal profile of the CW control field at different normalized propagation length, $\eta_pz/\gamma$. In Fig. \ref{fig:effective_control_rabi_frequency_at_different_z}, the continuous control field amplitude depletes with increasing propagation length as a consequence of the decay of population from state $|3\rangle$ to $|1\rangle$. Also its envelop gets distorted inwards in the shape of the Gaussian probe pulse due to the interdependence of effective control Rabi frequency on the time profile of the probe as seen in Eq. \eqref{eq:efffective_control_rabi_frequency}. The probe amplification can be understood by considering the population distribution in the three states. Figure \ref{fig:population_distribution} shows the population distribution in the states, $|1\rangle$, $|2\rangle$, and $|3\rangle$, respectively at the normalized propagation length $\eta_pz/\gamma = 5$. Unlike an usual three level $\Lambda$ system in atomic vapour, the $|1\rangle\leftrightarrow|3\rangle$ transition in Fig. \ref{fig:level_system} is not dipole forbidden for a polar molecular system, and the interaction strength of the third field with the induced dipole moment for $|1\rangle\leftrightarrow|3\rangle$ transition is high enough to cause population transfer from $|1\rangle$ to $|3\rangle$. The population from $|3\rangle$ is then excited to the state $|2\rangle$ due to the presence of strong control field in $|3\rangle\leftrightarrow|2\rangle$ channel. The nonzero population in the excited state, $|2\rangle$ can then decay to state $|1\rangle$ via stimulated emission in presence of the probe pulse, causing probe amplification. The nonzero population in the excited state can be seen in Fig. \ref{fig:population_distribution}.

\begin{figure}[!t]
\begin{center}
\includegraphics[width=\linewidth]{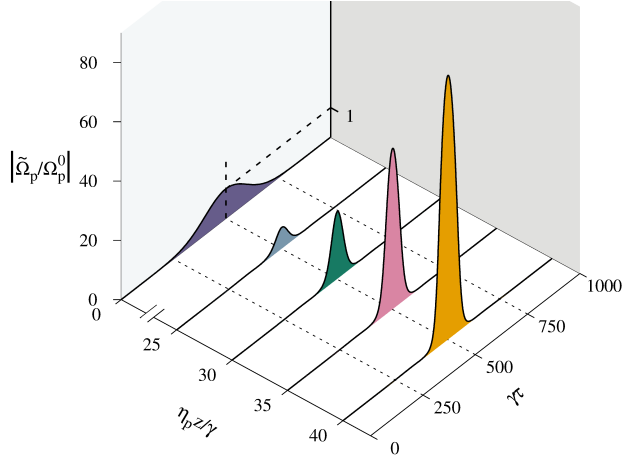}
\end{center}
\caption{\label{fig:effective_probe_rabi_frequency_at_different_z_two_photon}$\left|\tilde{\Omega}_p/\Omega^0_p\right|$ vs $\gamma \tau$ at different normalized propagation length, $\eta_pz/\gamma$. Parameters used: $n_p= 1, n_c=n_t = 2$, meaning probe transition is one photon, while rest two are two photon. All parameters remain same as Fig. \ref{fig:effective_probe_rabi_frequency_at_different_z}, except control field Rabi frequency at $z = 0$ is taken to be, $\Omega_c = 2\gamma$. The right and left $z$ axes represent normalized probe field magnitude at $z = 0$ and $z > 0$, respectively.} 
\end{figure}

Next we consider the case where the control and the third field transitions are two photon transitions while the probe transition is one photon, i.e., $n_p = 1$, $n_c = n_t = 2$. Figure \ref{fig:effective_probe_rabi_frequency_at_different_z_two_photon} shows the temporal profile of the probe pulse at different normalized propagation length, $\eta_pz/\gamma$. In Fig. \ref{fig:effective_probe_rabi_frequency_at_different_z_two_photon}, the probe pulse is amplified with increasing propagation length due to the same population redistribution process mentioned earlier in Fig. \ref{fig:population_distribution}. However, here the probe field has a frequency of $\omega_{21}$ while the control field has a frequency, $\omega_{23}/2$ which is close to half of the probe frequency from table \ref{table:parameters}. Thus in presence of PDM, due to unprohibited two photon excitation, it is possible to amplify a probe signal with the help of another signal whose frequency is half of the probe signal's frequency.
%
%
\section{\label{sec:level4}Conclusion}
In conclusion, we have investigated the propagation of a weak Gaussian probe pulse through a closed three level $\Lambda$ with permanent dipole moments (PDM) in presence of a strong continuous wave control field and a third field. We observed that presence of PDMs give rise to effective Rabi frequencies inside the medium which show oscillatory behavior with respect to their corresponding usual Rabi frequencies. In presence of PDMs all phenomenons are required to be explained in terms of the effective Rabi frequencies instead of the usual Rabi frequencies to get correct results. The presence of PDMs enables multiphoton excitation for the probe, control, and the third field transitions which is not possible in absence of PDMs. This allows for the possibility to amplify a probe signal, with a frequency which is twice of the control field.
%
%
\section{Reference}
\bibliography{references}

\providecommand{\newblock}{}
\begin{thebibliography}{10}
\expandafter\ifx\csname url\endcsname\relax
  \def\url#1{{\tt #1}}\fi
\expandafter\ifx\csname urlprefix\endcsname\relax\def\urlprefix{URL }\fi
\providecommand{\eprint}[2][]{\url{#2}}

\bibitem{eit_1}
Boller K~J, Imamo\ifmmode~\breve{g}\else \u{g}\fi{}lu A and Harris S~E 1991
  {\em Phys. Rev. Lett.\/} {\bf 66}(20) 2593--2596

\bibitem{eit_2}
Fleischhauer M, Imamoglu A and Marangos J~P 2005 {\em Rev. Mod. Phys.\/} {\bf
  77}(2) 633--673

\bibitem{slow_light_1}
Kasapi A, Jain M, Yin G~Y and Harris S~E 1995 {\em Phys. Rev. Lett.\/} {\bf
  74}(13) 2447--2450

\bibitem{slow_light_2}
Schmidt O, Wynands R, Hussein Z and Meschede D 1996 {\em Phys. Rev. A\/} {\bf
  53}(1) R27--R30

\bibitem{slow_light_3}
Kash M~M, Sautenkov V~A, Zibrov A~S, Hollberg L, Welch G~R, Lukin M~D,
  Rostovtsev Y, Fry E~S and Scully M~O 1999 {\em Phys. Rev. Lett.\/} {\bf
  82}(26) 5229--5232

\bibitem{storage_and_retrieval_1}
Dey T~N and Agarwal G~S 2003 {\em Phys. Rev. A\/} {\bf 67}(3) 033813

\bibitem{storage_and_retrieval_2}
Phillips D~F, Fleischhauer A, Mair A, Walsworth R~L and Lukin M~D 2001 {\em
  Phys. Rev. Lett.\/} {\bf 86}(5) 783--786

\bibitem{storage_and_retrieval_3}
Phillips N~B, Gorshkov A~V and Novikova I 2011 {\em Phys. Rev. A\/} {\bf 83}(6)
  063823

\bibitem{storage_and_retrieval_4}
Zhang S, Zhou S, Loy M~M~T, Wong G~K~L and Du S 2011 {\em Opt. Lett.\/} {\bf
  36} 4530--4532

\bibitem{stirap_1}
Bergmann K, N\"{a}gerl H~C, Panda C, Gabrielse G, Miloglyadov E, Quack M,
  Seyfang G, Wichmann G, Ospelkaus S, Kuhn A, Longhi S, Szameit A, Pirro P,
  Hillebrands B, Zhu X~F, Zhu J, Drewsen M, Hensinger W~K, Weidt S, Halfmann T,
  Wang H~L, Paraoanu G~S, Vitanov N~V, Mompart J, Busch T, Barnum T~J, Grimes
  D~D, Field R~W, Raizen M~G, Narevicius E, Auzinsh M, Budker D, P{\'{a}}lffy A
  and Keitel C~H 2019 {\em Journal of Physics B: Atomic, Molecular and Optical
  Physics\/} {\bf 52} 202001

\bibitem{stirap_2}
Shore B~W 2017 {\em Adv. Opt. Photon.\/} {\bf 9} 563--719

\bibitem{stirap_3}
Vitanov N~V, Rangelov A~A, Shore B~W and Bergmann K 2017 {\em Rev. Mod.
  Phys.\/} {\bf 89}(1) 015006

\bibitem{LWI_1}
Imamolu A, Field J~E and Harris S~E 1991 {\em Phys. Rev. Lett.\/} {\bf 66}(9)
  1154--1156

\bibitem{LWI_2}
Agarwal G~S 1991 {\em Phys. Rev. A\/} {\bf 44}(1) R28--R30

\bibitem{LWI_3}
Zhu Y 1992 {\em Phys. Rev. A\/} {\bf 45}(9) R6149--R6152

\bibitem{eit_pdm_1}
Zhou F, Niu Y and Gong S 2009 {\em The Journal of Chemical Physics\/} {\bf 131}
  034105

\bibitem{eit_pdm_2}
Singh N, Lawande Q~V, D~Souza R and Jagatap B~N 2012 {\em The Journal of
  Chemical Physics\/} {\bf 137} 104309

\bibitem{stirap_pdm_1}
Brown A 2007 {\em Chemical Physics\/} {\bf 342} 16--24 ISSN 0301-0104

\bibitem{stirap_pdm_2}
Deng L, Niu Y and Gong S 2018 {\em Phys. Rev. A\/} {\bf 98}(6) 063830

\bibitem{propagation_pdm_1}
Du Q, Hang C and Huang G 2014 {\em J. Opt. Soc. Am. B\/} {\bf 31} 594--602

\bibitem{higher_harmonic_generation_pdm_1}
Calderon O, Gutierrez-Castrejon R and Guerra J 1999 {\em IEEE Journal of
  Quantum Electronics\/} {\bf 35} 47--52

\bibitem{pop_inversion_pdm_1}
Macovei M, Mishra M and Keitel C~H 2015 {\em Phys. Rev. A\/} {\bf 92}(1) 013846

\bibitem{optical_bistability_pdm_1}
Asadpour S~H and Soleimani H~R 2014 {\em J. Opt. Soc. Am. B\/} {\bf 31}
  3123--3130

\bibitem{unitary_transformation}
Meath W~J and Power E~A 1984 {\em Molecular Physics\/} {\bf 51} 585--600

\bibitem{HCN_data_1}
Bowman J~M, Irle S, Morokuma K and Wodtke A 2001 {\em The Journal of Chemical
  Physics\/} {\bf 114} 7923--7934

\bibitem{HCN_data_2}
 2002 {\em Spectrochimica Acta Part A: Molecular and Biomolecular
  Spectroscopy\/} {\bf 58} 673--690 ISSN 1386-1425

\bibitem{decay_rate}
Lavoine J~P 2007 {\em The Journal of Chemical Physics\/} {\bf 127} 094107

\end{thebibliography}
\end{document}